\documentclass[lettersize,journal]{IEEEtran}
\usepackage{amsmath,amsfonts}
\usepackage{algorithmic}
\usepackage{algorithm}
\usepackage{array}
\usepackage[caption=false,font=normalsize,labelfont=sf,textfont=sf]{subfig}
\usepackage{textcomp}
\usepackage{stfloats}
\usepackage{url}
\usepackage{verbatim}
\usepackage{graphicx}
\usepackage{cite}
\usepackage{xcolor}
\usepackage{tikz}
\usepackage{multirow}
\begin{document}

\title{Node Placement and Path Planning for Improved Area Coverage in Mixed Wireless Sensor Networks}

\author{Survi Kumari and
Seshan Srirangarajan,~\IEEEmembership{Member,~IEEE}}



\maketitle

\begin{abstract}
For the large-scale monitoring of a physical phenomena using a wireless sensor network (WSN), a large number of static and/or mobile sensor nodes are required, resulting in higher deployment cost. In this work, we develop an efficient algorithm that can employ a small number of static nodes together with a set of mobile nodes for improved area coverage. An efficient deployment of static nodes and guided mobility of the mobile nodes is critical for maximizing the area coverage. To this end, we propose three mixed integer linear programming (MILP) formulations. The first formulation efficiently deploys a set of static nodes and the other two formulations plan the path of a set of mobile nodes so as to maximize the area coverage and minimize the total number of movements required to achieve the desired coverage. We present extensive performance evaluation of the proposed algorithms and its comparison with benchmark approaches. The simulation results demonstrate the superior performance of the proposed algorithms for different network sizes and number of static and mobile nodes.
\end{abstract}


\section{Introduction}
Wireless sensor networks (WSN) are being used in various monitoring or surveillance applications and ensuring full area coverage is one of the key objectives in such applications~\cite{Savkin2019}. Deployment of only static nodes typically leads to coverage holes and overlapping coverage due to sub-optimal placement of nodes and/or nodes becoming non-functional over a period of time after the initial deployment. Increasing the number of nodes or their sensing region are typically not cost effective. In addition, these measures do not address the issue of non-functional nodes or overlapping coverage or the effect of environmental factors on the network. Thus, a mixed WSN, a network with combination of static and mobile nodes, has been proposed to address these limitations~\cite{Lambrou2013,Zhu_Fan_Wu_Wen_2016}.

Mobility in sensor nodes is the ability of the nodes to move and change their locations post their initial deployment. With the advancement in technology, the applications of mobile sensor nodes have gradually increased. The use of mobile nodes significantly improves the possibility of maintaining a robust network coverage~\cite{Liu_mobilty2005}. In WSN literature and applications, the mobility of nodes is primarily exploited in two ways. In some cases, the fusion center or sink is mobile and the mobile sink moves throughout the network area to gather data from the sensing nodes~\cite{Liang2010,Zhong2018}. In other cases, one or more of the sensing nodes are mobile and move through the network area to record data along with location information and deliver the collected data to the fusion center.

A WSN is usually deployed in an area of interest for monitoring or detection applications. The ability to monitor the entire area of interest at all time instances is referred to as continuous coverage. However, in many applications, periodic monitoring is sufficient instead of continuous monitoring, and this is referred to as sweep coverage. Typical examples include data gathering and message ferrying~\cite{Xiaofeng2022,Gao2016}. 
For a mixed WSN, efficient path planning strategies need to be developed for mobile nodes to achieve sweep/periodic coverage. 

Coverage path planning has been and continues to be an active area of research with the growing popularity of mobile nodes in the form of mobile robots and unmanned aerial vehicles. The use of mobile nodes in WSNs for improving sweep coverage has been addressed through different approaches. The authors in~\cite{GAO2018} presented approximation algorithms that minimize the time taken by mobile nodes for visiting a set of targets. The objective is to reduce the target detection period, while minimizing the trajectory length of the mobile nodes. In~\cite{BiddingP2007}, the static sensors detect coverage holes by using Voronoi diagrams and bid for mobile sensors based on the size of the coverage hole detected by them. Mobile sensors accept the highest bids and move to heal the largest holes. To reduce the movement distance of mobile nodes, a proxy-based bidding protocol is employed where mobile sensors perform virtual movements from small holes to large holes and only perform physical movements after the final destinations are identified.

In~\cite{Lambrou2013}, the authors consider various cost measures to address trade-offs between area coverage and distance traveled by mobile nodes or information exchange among nodes. In~\cite{VECCHIO2015}, the authors proposed a distributed technique for iteratively computing the paths for mobile nodes in a greedy manner. They employ a bidding approach similar to that in~\cite{BiddingP2007} while using the movement scheme described in~\cite{DPP2010, Lambrou2013}. In addition, the static nodes employ the zoom algorithm to determine the largest hole~\cite{LAMBROU2007,Lambrou2009}. A review of several other methods that exploit mobility of nodes for network coverage problems can be found in~\cite{Elhabyan2019,TEMENE2022}.

Our work is inspired by~\cite{ZYGOWSKI2020}, which proposes an optimization framework to plan the path of a single mobile node to maximize area coverage and/or minimize the total trip time for the mobile node. In this work, we present strategies to plan the path of a set of mobile nodes to achieve sweep coverage through periodic monitoring of the network area. The proposed system model assumes a small number of mobile sensor nodes along with a limited number of static sensor nodes. The objective is to plan the paths of the mobile nodes, primarily over the uncovered areas, so as to improve the overall area coverage and total trip time. We also propose an algorithm for the deployment of static nodes that aids in the path planning of the mobile nodes by avoiding the challenges that can arise due to a random deployment of static nodes, such as boundary coverage holes~\cite{WATFA2007}, network partition, and redundant coverage~\cite{DPP2010}. The proposed system model includes a parameter that determines the permissible level of redundant coverage or overlapping coverage during the path planning process. The proposed strategies are formulated as mixed integer linear programming (MILP) problems. The key contributions of this work are as follows:
\begin{itemize}
\item An MILP-Static formulation that places static nodes to maximize area coverage with more weight given to covering the network boundary areas.
\item An MILP-Cov formulation that plans the path of a set of mobile nodes while maximizing the coverage area within a given number of movements or time steps.
\item An MILP-Mov formulation that plans the path of a set of mobile nodes while minimizing the number of movements required to achieve a desired coverage level.
\item Extensive simulations comparing performance of the proposed algorithms with benchmark methods and computational complexity analysis.
\end{itemize}
%
\section{System Model}
Consider a rectangular network area which is subdivided into grids with ${M} \times {N}$ unit square cells. Each grid cell location is denoted by coordinates $(i,j)$ with $1 \leq i \leq M$ and $1 \leq j \leq N$. We assume a mixed WSN, consisting of a few static and mobile nodes for monitoring the network area, and a single data sink node. 
The first objective is to place the static nodes within the network area in a manner that aids the path planning of the mobile nodes. Next, we plan the paths of the mobile nodes to maximize the area coverage and minimize the trip time. The key parameters of the proposed system model are listed in Table~\ref{Table:notations}, which are described next.
\begin{figure}
\centering
\includegraphics[width= 0.7\linewidth]{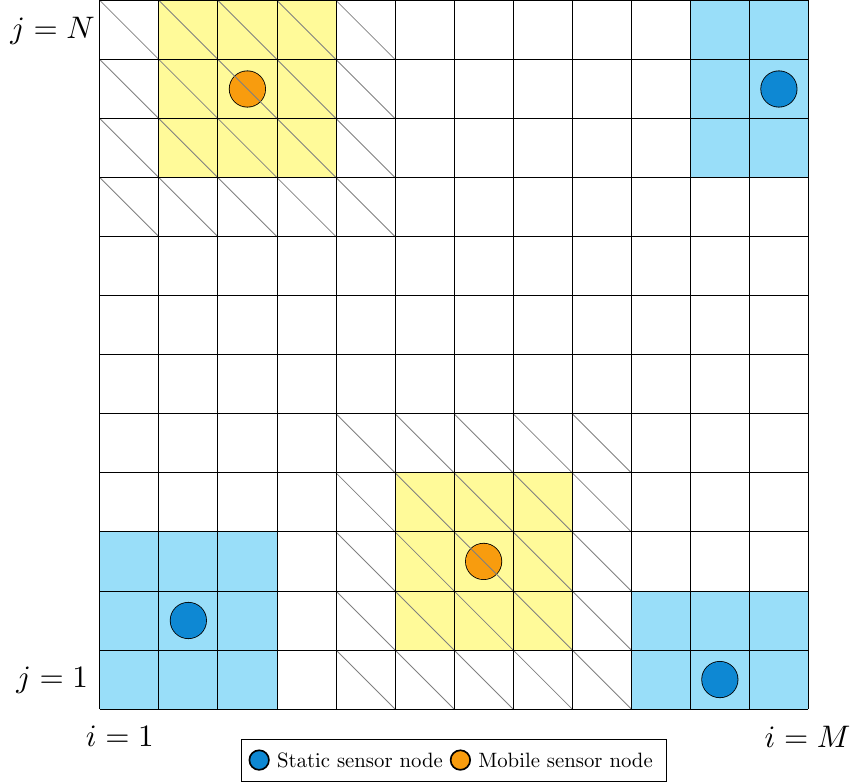}
\caption{A typical network area. 
The colored cells around a sensor node indicate the sensing region of that sensor node ($r_s=1$). The cells marked with diagonal lines around each mobile node represent potential locations where the mobile node could move to in the next iteration ($\rho_x=\rho_y=2$).}
\label{fig:network_model}
\end{figure}
\begin{table}
\caption{Notations}
\label{table_example}
\begin{tabular}{|c|p{7cm}|}
\hline
\bfseries  Symbol  & \bfseries  Definition \\
\hline
$N_{s}$ & Number of static sensor nodes\\
$L$ & Number of mobile sensor nodes\\
$\mathcal{C}$, $|\mathcal{C}|$ & Set of all the cells in the network area and its cardinality\\
$\mathcal{B}$ & Set of cells at the boundary of the network area\\
$\mathcal{A}$ & Set of cells other than the boundary cells\\
$\mathcal{C}_1$ & Set of cells not covered by static nodes\\
$\mathcal{C}_2$ & Set of cells covered by static nodes\\
$cr$ & Area coverage ratio\\
$r_{s}$ & Sensing radius of a sensor node\\
$\rho_{x}$, $\rho_{y}$ & One-step traveling range of mobile node along $x$ and $y$ directions\\
$c_o$ & Overlapping cell coverage factor\\
$K_{\text{max}}$ & Maximum number of iterations\\
\hline
$x^{s}_{i,j}$ & $x^{s}_{i,j}=1$ if the $s^{\text{th}}$ static sensor node is located at cell $(i,j)$, otherwise $x^{s}_{i,j}=0$.\\
$x^{l,k}_{i,j}$ & $x^{l,k}_{i,j}=1$ if the $l^{\text{th}}$ mobile node is located at cell $(i,j)$ in the $k^{\text{th}}$ iteration.\\
\hline
$c^{s}_{i,j}$ & $c^{s}_{i,j}=1$ if the cell $(i,j)$ is covered by the $s^{\text{th}}$ static sensor node, otherwise $c^{s}_{i,j}=0$.\\
$c^{l,k}_{i,j}$ & $c^{l,k}_{i,j}=1$ if the cell $(i,j)$ is covered by the $l^{\text{th}}$ mobile node in the $k^{\text{th}}$ iteration, otherwise $c^{l,k}_{i,j}=0$.\\
$c_{i,j}$ & $c_{i,j}=1$ if the cell $(i,j)$ is covered by a sensor node during any iteration, otherwise $c_{i,j}=0$.\\
\hline
\end{tabular}
\label{Table:notations}
\end{table}

Let $N_{s}$ and $L$ denote the number of static and mobile nodes, respectively. The coordinate of the grid cell in which a sensor is located is defined as the location of that sensor. Let $r_{s}$ denote the sensing radius (or sensing range) of static/mobile nodes such that $r_{s}$ is the number of cells that can be sensed by a node in each direction from its current location. The network is assumed to be connected such that the sensor nodes can communicate with the sink node at all times. 

After the initial deployment of static nodes is completed, the paths to be followed by the mobile nodes is planned centrally. The path planning is an iterative procedure where the location of each mobile node is computed simultaneously at each iteration. The mobile nodes sense the parameter(s) of interest in the cells within their sensing range $r_{s}$ and then move to their next locations within the one-step traveling range, denoted by $\rho_{x}$ and $\rho_{y}$ along the $x$ and $y$ directions, respectively. The one-step traveling range of a mobile node is the maximum number of cells up to which the mobile node can move in each direction from its current location in one iteration. The coverage ratio, denoted by ${cr}$, is the ratio of the number of grid cells covered at least once in any iteration and the total number of grid cells within the network area; thus, $0 \leq {cr} \leq 1$. An example network area is shown in Fig.~\ref{fig:network_model}. 
%
\section{Proposed Strategies}
\label{sec:proposed_strategy}
In this section, we describe the proposed algorithms for the placement of static nodes and path planning of the mobile nodes. Consider the variables $x_{i,j}^{s}$ and $x_{i,j}^{l,k}$ which indicate the locations of static and mobile nodes, respectively. Also, let $c_{i,j}^{s}$, $c_{i,j}^{l,k}$, and $c_{i,j}$ be the coverage variables. These variables are defined in Table~\ref{Table:notations}. We define $x^{s}_{i,j}$ and $x^{l,k}_{i,j}$ as binary variables, whereas $c_{i,j}$, $c^{s}_{i,j}$, and $c^{l,k}_{i,j}$ are defined as continuous variables in the range $\left[0,1\right]$. Thus, the proposed node placement and path planning strategies result in MILP formulations which are described next.
\subsection{Static Node Placement}
\begin{figure}
\centering
\includegraphics[width= 0.9\linewidth]{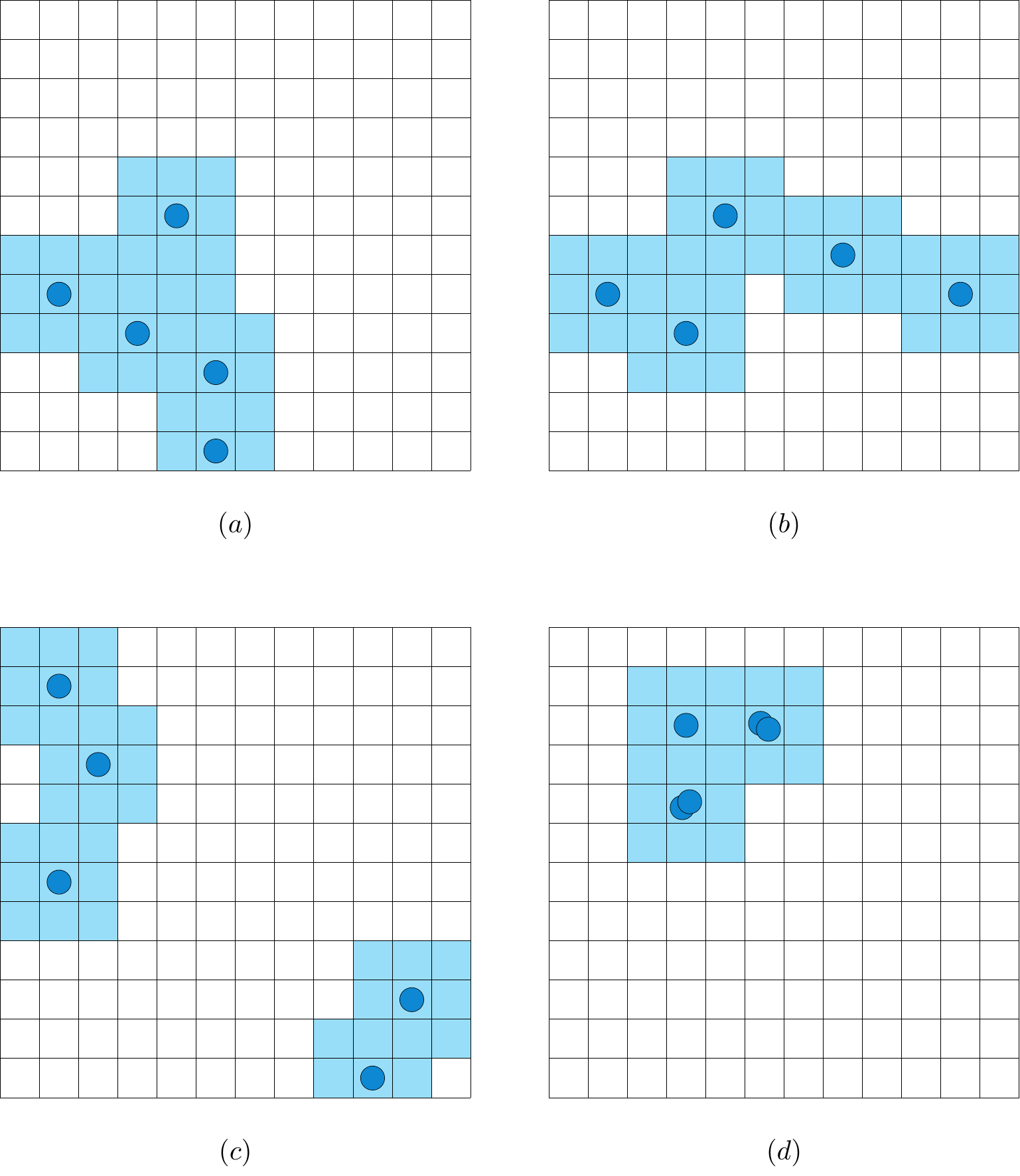}
\caption{Possible challenges due to random deployment of static sensor nodes in a network area. $(a)$, $(b)$ network partition, $(c)$ boundary coverage hole, and $(d)$ overlapping/redundant coverage.}
\label{fig:random_deployment_scene}
\end{figure}

The first step is to place the static sensor nodes within the network area so as to maximize the area coverage. As noted earlier, the random deployment of static nodes can result in challenges such boundary coverage holes, network partitioning, and redundant coverage. Some of these scenarios are illustrated in Fig.~\ref{fig:random_deployment_scene}. In addition, if the static nodes do not cover the boundary cells, the mobile nodes would have to cover these resulting in a larger number of movements and redundant coverage. Thus, in the proposed static node placement strategy we would like to maximize the covered area while giving more importance/weight to covering the cells at the network boundary. 
%

Let $\mathcal{B}$ represent the set of cells at the boundary of the network area. It can be seen that $|\mathcal{B}|=2(M+N-2)$. The static node placement can be formulated as the following optimization problem, referred to as MILP-Static. 
%
%
\begin{align}
    & \max \left(\sum_{\substack{s=1\\(i,j)\in \mathcal{A}}}^{N_s}{c^{s}_{i,j}} + \sum_{\substack{s=1\\(i,j)\in \mathcal{B}}}^{N_s}{ \alpha c^{s}_{i,j}}\right)\label{eq:static_obj}\\
    & \text{s.t.} \sum_{(i,j)\in \mathcal{C}}{x^{s}_{i,j}} = 1, \quad s = 1,\dots,N_{s}\label{eq:static_pos_constraint}\\
    & {{c^{s}_{i,j}} = \sum_{p=-r_{s}}^{r_{s}} \sum_{q=-r_{s}}^{r_{s}} x^{s}_{i+p,j+q}}, \forall (i,j) \in \mathcal{C}_{1}, s = 1,\dots,N_{s}\label{eq:static_cov_constraint}\\
    & \sum_{s=1}^{N_{s}} {c^{s}_{i,j}} \leq c_o, \quad \forall (i,j) \in \mathcal{C}\label{eq:static_overlap_constraint}
\end{align}
%
%
%

The objective function in~\eqref{eq:static_obj} maximizes the total number of covered cells, where $\alpha$ is a weight parameter associated with the coverage of the boundary cells. Using $\alpha > 1$, gives more importance to the coverage of the boundary cells. We use $\alpha=4$ in our simulations. \eqref{eq:static_pos_constraint} is the position constraint which ensures that each static node is placed in only one cell. The constraint~\eqref{eq:static_cov_constraint} ensures that a cell $(i,j)$ is considered as covered by the $s^{\text{th}}$ static node, i.e., $c^{s}_{i,j} = 1$, if the cell $(i,j)$ is within the sensing range of the $s^{\text{th}}$ static node. The overlapping coverage constraint~\eqref{eq:static_overlap_constraint} ensures that each cell is covered by at most $c_o$ static nodes.
%
\subsection{Coverage Maximization}
After the static nodes have been placed (using MILP-Static), the mobile nodes must traverse the network area to cover the uncovered cells $(i,j) \in \mathcal{C}_{1}$, so as to maximize the area coverage. The path planning of the mobile nodes, to maximize area coverage, can be formulated as the following optimization problem and is referred to as MILP-Cov.
\begin{align}
    &\max \sum_{(i,j)\in \mathcal{C}_{1}}{c_{i,j}}\label{eq:cov_obj}\\
    & \text{s.t.} \sum_{(i,j)\in \mathcal{C}_{1}}{x^{l,k}_{i,j}} = 1, \quad l= 1,\dots,L, k = 1,\dots,K_{\text{max}}\label{eq:cov_pos}\\
    & x^{l,k+1}_{i,j} = \sum_{p=-\rho_{x}}^{\rho_{x}} \sum_{q=-\rho_{y}}^{\rho_{y}}{x^{l,k}_{i+p,j+q}}, \quad\forall (i,j) \in \mathcal{C}_{1},\label{eq:cov_mobility}\\
    & \quad\quad\quad\quad\quad\quad\quad l= 1,\dots,L, k = 1,\dots,(K_{\text{max}}-1)\nonumber\\
    & c^{l,k}_{i,j} = \sum_{p=-r_{s}}^{r_{s}} \sum_{q=-r_{s}}^{r_{s}}{x^{l,k}_{i+p,j+q}}, \quad \forall (i,j) \in \mathcal{C}_{1},\label{eq:cov_cell_1}\\
    & \quad\quad\quad\quad\quad\quad\quad l= 1,\dots,L, k = 1,\dots,K_{\text{max}}\nonumber\\
    & {c_{i,j}} \geq {c^{l,k}_{i,j}},\forall (i,j) \in \mathcal{C}_{1},\; \forall l= 1,\dots,L, k = 1,\dots,K_{\text{max}}\label{eq:cov_cell_2}\\
    & c_{i,j} \leq \sum_{l=1}^{L} \sum_{k=1}^{K_{\text{max}}}{c^{l,k}_{i,j}}, \quad \forall (i,j) \in \mathcal{C}_{1}\label{eq:cov_cell_3}\\
    & \sum_{l=1}^{L} \sum_{k=1}^{K_{\text{max}}} {c^{l,k}_{i,j}} \leq c_o, \quad \forall (i,j) \in \mathcal{C}_{1}\label{eq:cov_overlap}\\
    & 0 \leq c_{i,j}, c_{i,j}^{l,k} \leq 1, \forall (i,j) \in \mathcal{C}_1, l=1,\dots,L, k=1,\dots,K_{\text{max}}\label{eq:c_range} 
\end{align}
%

The objective function in~\eqref{eq:cov_obj} maximizes the area covered by the mobile nodes from the uncovered set as they traverse the network area over $K_{\text{max}}$ iterations. \eqref{eq:cov_pos} is the position constraint which ensures that each mobile node is located in only one cell in any given iteration. The mobility constraint in~\eqref{eq:cov_mobility} restricts the distance that each mobile node can travel along $x$ and $y$ directions to $\rho_{x}$ and $\rho_{y}$, respectively. \eqref{eq:cov_cell_1}-\eqref{eq:cov_cell_3} represent the cell coverage constraints. Constraint~\eqref{eq:cov_cell_1} states that cell $(i,j)$ is said to be covered by the $l^{\text{th}}$ mobile node in the $k^{\text{th}}$ iteration, i.e., $c^{l,k}_{i,j} = 1$, if the cell $(i,j)$ is within the sensing range of the $l^{\text{th}}$ mobile node at the $k^{\text{th}}$ iteration. Constraint~\eqref{eq:cov_cell_2} sets $c_{i,j}$ to zero if cell $(i,j)$ is not covered by any mobile node during any iteration. Constraint~\eqref{eq:cov_cell_3} ensures that if a cell is covered in any iteration by a mobile node, it is considered as covered for the rest of the algorithm. Constraint~\eqref{eq:cov_overlap} allows for limited overlap or redundancy in coverage (i.e., allows each cell to be covered at most $c_o$ times) due to multiple mobile nodes covering the same cell across iterations. This is essential to ensure that area coverage can be maximized while allowing some redundancy in coverage. Setting $c_o=1$ may result in the mobile node paths getting stuck at a cell before $K_{\text{max}}$ iterations at the cost of limiting the area coverage.
%
\subsection{Movement Minimization}
An alternative strategy for planning the paths of the mobile nodes is to minimize the total number of movements by the mobile nodes for attaining a desired coverage ratio ($cr$). This can be formulated as the following optimization problem and is referred to as MILP-Mov.
\begin{align}
    \min & \sum_{k=1}^{K_{\text{max}}} \sum_{\substack{l=1\\(i,j) \in \mathcal{C}_{1}}}^{L} x^{l,k}_{i,j}\label{eq:mov_obj}\\
    \text{s.t.} & \sum_{(i,j)\in \mathcal{C}_{1}} x^{l,k}_{i,j} \leq 1, l= 1,\dots,L, k = 1,\dots,K_{\text{max}}\label{eq:mov_pos}\\
    & \sum_{(i,j) \in \mathcal{C}}{c_{i,j}} \geq {cr}\cdot |\mathcal{C}|\label{eq:mov_cov}\\
    & \text{Constraints}~\eqref{eq:cov_mobility},~\eqref{eq:cov_cell_1},~\eqref{eq:cov_cell_2},~\eqref{eq:cov_cell_3},~\eqref{eq:cov_overlap},~\eqref{eq:c_range}\nonumber
\end{align}
%

The objective function in~\eqref{eq:mov_obj} minimizes the total number of cells that are visited by the mobile nodes while achieving the desired coverage ratio. The number of cells visited by the mobile nodes represents the number of mobile node movements. 
\eqref{eq:mov_pos} is the position constraint, which is similar to the position constraint~\eqref{eq:cov_pos} of MILP-Cov, except that MILP-Mov allows the possibility that after a certain number of iterations ${x^{l,k}_{i,j}}$ for all the mobile nodes can be zero. This implies that all the mobile nodes would stop their movements as soon as the desired coverage ratio is achieved. \eqref{eq:mov_cov} indicates that the total coverage achieved by the static and mobile nodes must satisfy the desired coverage ratio.

It is to be noted that even though $c_{i,j}$, $c_{i,j}^{s}$, and $c_{i,j}^{l,k}$ have been defined as continuous variables, due to their defined range and other constraints, they will behave as binary variables. This is because the constraints have been formulated such that they can only take values $0$ or $1$. They have been defined as continuous variables in order to reduce the complexity of the integer linear programs (ILPs). It is well-known that the complexity of ILPs increase exponentially with the number of binary/integer variables \cite{ZYGOWSKI2020}.
%
\section{Simulation Results}
In this section, we present detailed performance analysis of the proposed MILP-based node placement and path planning algorithms.
\subsection{Simulation Setup}
We consider network area of sizes ranging from $8 \times 8$ to $15 \times 15$ with grid cells of size $1 \times 1$. The number of static and mobile nodes are varied in the range $0$-$10$ and $1$-$5$, respectively. We assume sensing range $r_s = 1$, one-step traveling range $\rho_{x}=\rho_{y}=2$, and overlapping cell coverage factor $c_o = 3$ for MILP-Cov and MILP-Mov while $c_o =1$ for MILP-Static. The algorithms were implemented in MATLAB R2020b and used 12.10.0 version of IBM ILOG CPLEX optimization software. The CPLEX parameter settings used include `TimeLimit' of $18000$~s, `MIPGap' of $0$, and branch and cut method as the `MIP Strategy Search'.
%
%
\subsection{Static Node Placement}
We first compare the two static node placement strategies, random and MILP-Static, in terms of area coverage. We assume that after the static node placement, the paths of the mobile nodes are planned using MILP-Cov. Table~\ref{Random v/s MILP-static} compares the network area coverage achieved using different numbers of mobile and static sensor nodes for two network sizes. In the random placement strategy, the static nodes are placed according to a uniform distribution. The area coverage values reported are obtained with $K_{\text{max}}=4$ and by averaging over five simulation runs. The results indicate that the area coverage performance is significantly better in the case where the static nodes are placed using the MILP-Static strategy, as compared to random placement. Based on these results, for the rest of the simulations of MILP-Cov and MILP-Mov, the MILP-Static algorithm will be used for the placement of static nodes.
\begin{table}
\renewcommand{\arraystretch}{1.1} 
\caption{Area coverage (MILP-Cov) using two different static node placement strategies: Random and MILP-Static.}
\label{Random v/s MILP-static}
\centering
\begin{tabular}{|c|c|c|c|c|}
\hline
\multirow{2}{*}{Network Size} & \multirow{2}{*}{$L$} & \multirow{2}{*}{$N_{s}$} & \multicolumn{2}{c|}{Area Coverage ($\%$)}\\
\cline{4-5}
& & & Random & MILP-Static\\
\hline
\multirow{6}{*}{$8 \times 8$} & 1 & 3 & 63.18 & 85.93\\
& 1 & 5 & 68.23 & 100\\        
& 2 & 3  & 95.93 & 100\\
& 2 & 5  & 93.39 & 100\\       
& 3 & 3  & 100 & 100\\
& 3 & 5  & 94.47 & 100\\
\hline
\multirow{6}{*}{$10 \times 10$} & 1 & 3  & 42.12 & 60\\
& 1 & 5 & 44.91 & 77\\
& 2 & 3 & 75.11 & 88\\
& 2 & 5 & 78.14 & 96\\
& 3 & 3 & 93.24 & 99\\
& 3 & 5 & 94.88 & 100\\    
\hline
\end{tabular}
\end{table}
\subsection{MILP-Cov and MILP-Mov Performance}
\begin{figure}
\centering
 \includegraphics[width= 0.7\linewidth, trim = 3cm 9cm 4cm 9cm, clip]{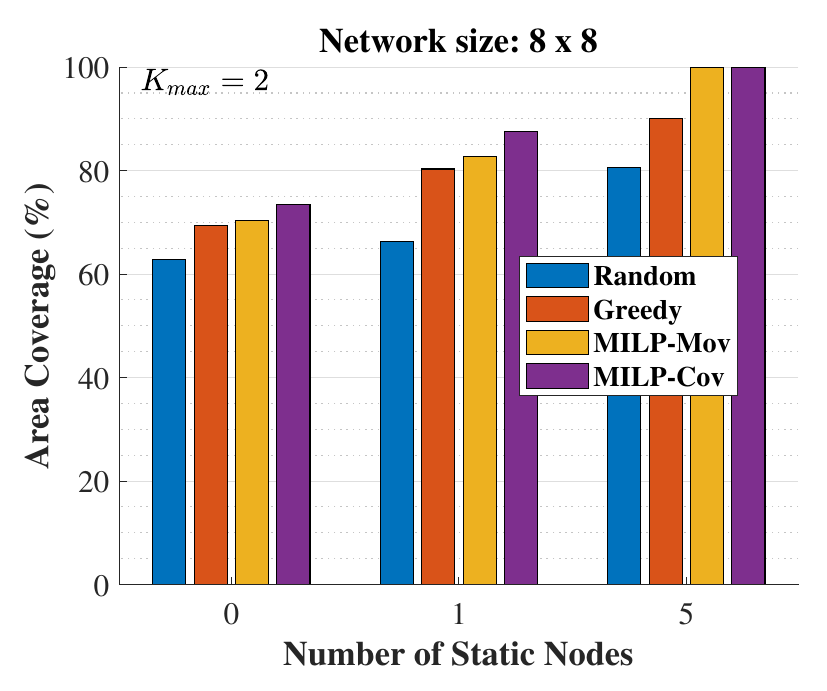}\\ (a)\\
 \includegraphics[width= 0.7\linewidth,trim = 3cm 9cm 4cm 9cm, clip]{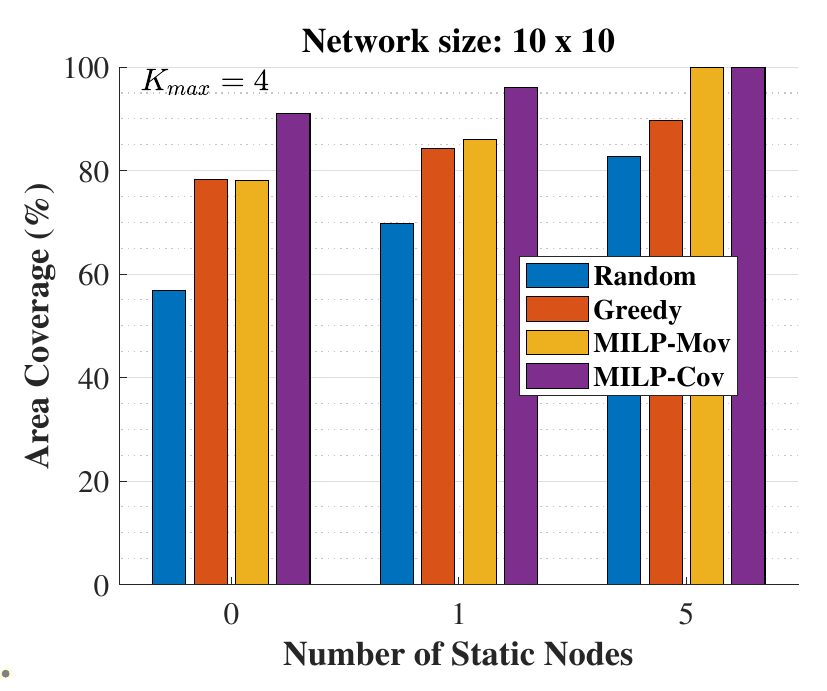}\\ (b)
\caption{Area coverage using three $(L=3)$ mobile nodes with different number of static nodes.} 
\label{fig:MILP_cov_perf}
\end{figure}

In Fig.~\ref{fig:MILP_cov_perf},
we compare the performance of MILP-Cov, MILP-Mov, random movement, and greedy approaches in terms of the area coverage, with three mobile nodes. In the greedy approach, each mobile node considers the cells within its one-step traveling range and moves to the cell location that would cover the maximum number of cells which are yet to be covered. On the other hand, in the random movement approach, each mobile node moves to a cell that is selected randomly from all the potential cells within its one-step traveling range. It is seen that MILP-Cov consistently outperforms random and greedy approaches with the greedy approach performing better than the random movement approach. With $N_{\text{s}} = 0 \text{ or }1$, the area coverage using MILP-Mov and greedy approaches is similar since with this level of coverage through static node, the overall coverage search space does not change significantly. 
%

\begin{figure}
\centering
 \includegraphics[width= 0.7\linewidth, trim = 3cm 9cm 4cm 9cm, clip]{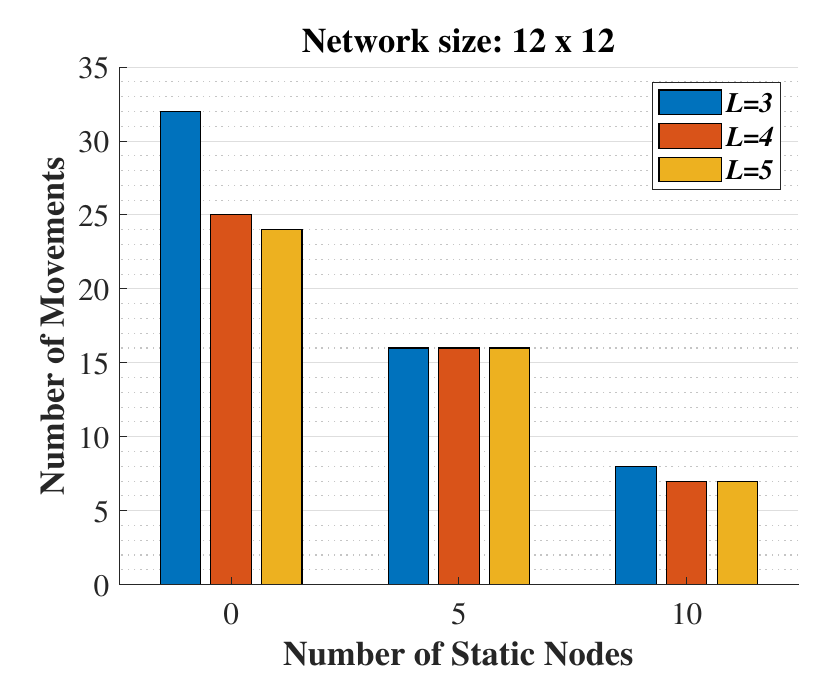}\\(a)\\
 \includegraphics[width= 0.7\linewidth, trim = 3cm 9cm 4cm 9cm, clip]{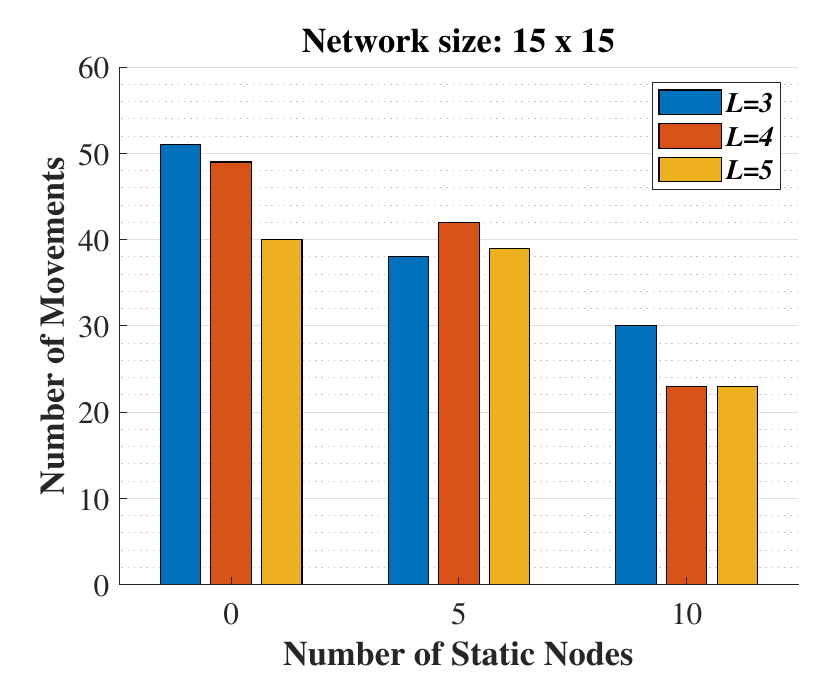}\\(b)
\caption{Number of movements required by mobile nodes for full area coverage $({cr = 1})$ using MILP-Mov with different number of static and mobile nodes.}
\label{fig:MILP_mov_perf}
\end{figure}

In Fig.~\ref{fig:MILP_mov_perf}, we consider the performance of the MILP-Mov strategy in terms of the number of movements needed to achieve full area coverage (${cr}=1$) for two different network sizes. The results show that, for a given network size, as the number of static and mobile nodes increases, the required number of movements decreases. 
%
%

\begin{figure}
\centering
\includegraphics[width= 0.85\linewidth,trim = 2cm 8cm 2cm 8cm, clip]{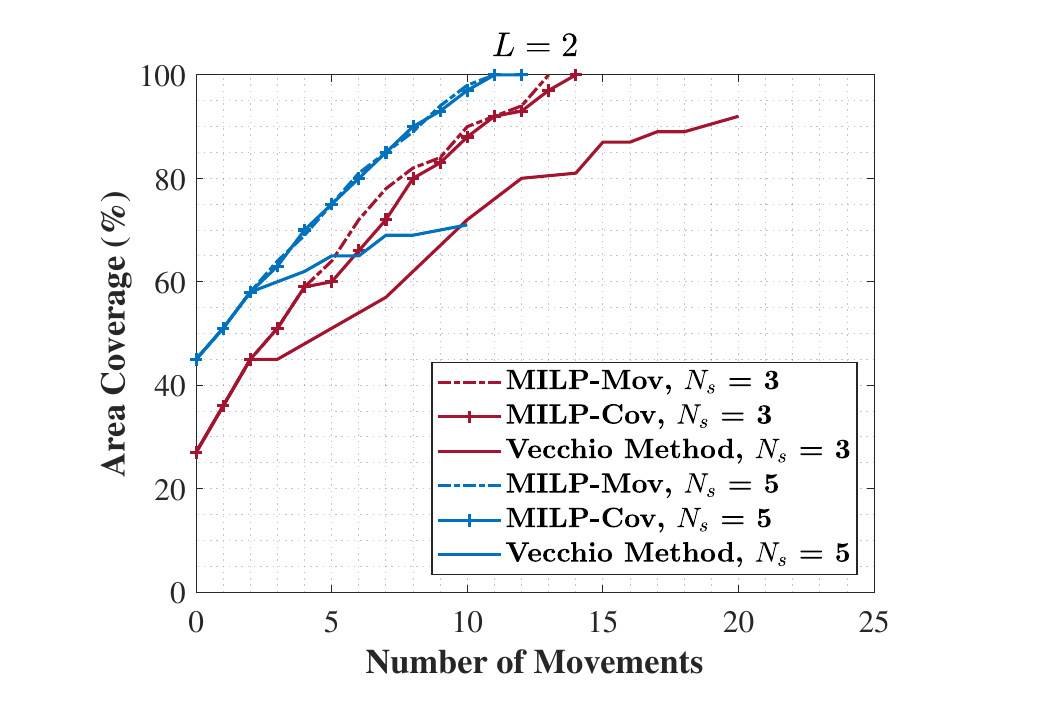} \\ (a) \\
\vspace{0.5cm}
\includegraphics[width= 0.85\linewidth,trim = 2cm 8cm 2cm 8cm, clip]{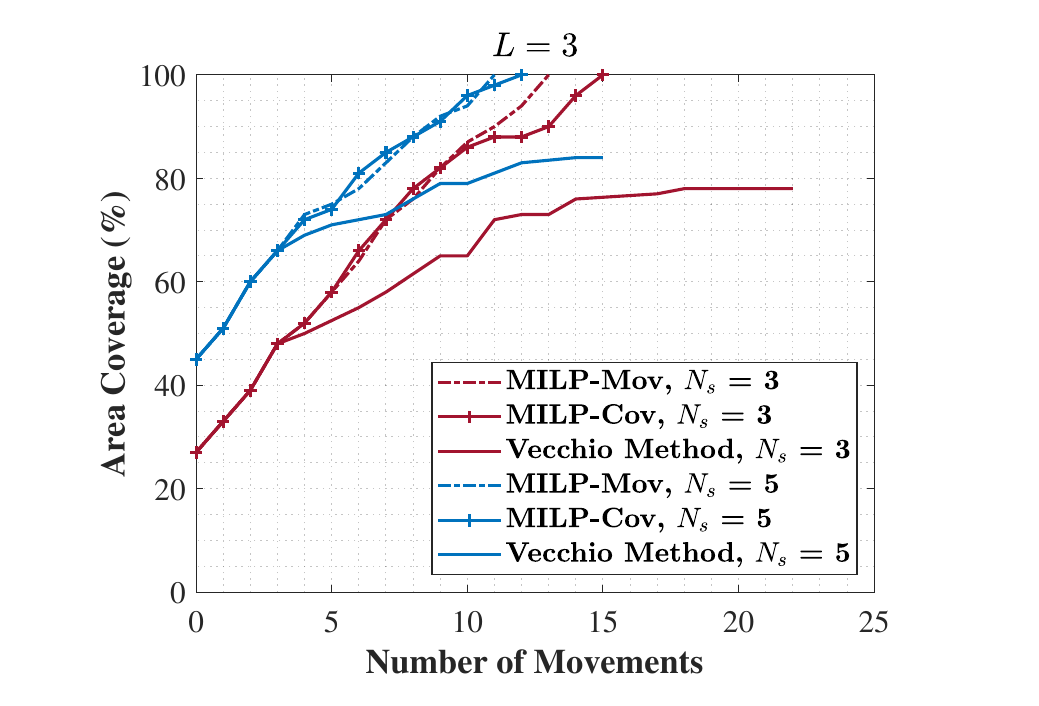} \\ (b)
\caption{Area coverage as a function of the number of movements by the mobile nodes (Network size = $10 \times 10$).}
\label{fig:area_cov_vs_movements}
\end{figure}

In Fig.~\ref{fig:area_cov_vs_movements}, we compare performance of the MILP-Cov, MILP-Mov algorithms, and the path planning technique presented in~\cite{VECCHIO2015}, in terms of the area coverage as a function of the number of mobile node movements. We refer to the method presented in~\cite{VECCHIO2015} as Vecchio method, based on the first author's last name. For a fair comparison, the initial locations of the static and mobile nodes are kept the same for all the three algorithms. The initial locations of static and mobile nodes are obtained based on MILP-Static and MILP-Mov algorithms, respectively.
For implementing the Vecchio method, we use $r_s = 1.5$, $r_c = 2r_s$, $\mu = 2$, $\rho = 2$, $\phi = \pi/6$, $n = 10$, and these parameters have the same definitions as in~\cite{VECCHIO2015}. The values of weight parameters associated with the cost function and other parameters are as given in~\cite{VECCHIO2015}. 
From Fig.~\ref{fig:area_cov_vs_movements}, it is seen that the proposed MILP-Cov and MILP-Mov algorithms achieve full area coverage, and outperform the Vecchio method~\cite{VECCHIO2015} by requiring much fewer mobile node movements for achieving a given level of area coverage.
It is observed that the Vecchio method does not achieve full area coverage even with a very large number of movements and the coverage saturates at some point. In addition, in the Vecchio method, the mobile nodes stop their movements before achieving full coverage as they are unable to locate the uncovered areas and wait endlessly while looking for the next location using the zoom algorithm~\cite{LAMBROU2007,Lambrou2009}. Although, the three methods show improved coverage performance with increase in the number of static and mobile nodes.
%
\subsection{Computational Complexity Analysis}
We analyze the complexity of the proposed methods in terms of the number of continuous variables, number of integer variables, and the number of constraints. These are listed in Table~\ref{complexity_table} and it is seen that these complexity measures increase linearly with the network size and the number of nodes in the network. 
In Table~\ref{Movements v/s Iterations},  we list the number of movements along with the computational time required for full coverage by MILP-Cov and MILP-Mov methods for different network sizes and with different number of static and mobile nodes.
The static node placement is performed using the proposed MILP-Static strategy. The simulations are performed on an Intel(R) Core(TM) i7-8550U CPU @ $1.80$~GHz-$1.99$~GHz, $16$~GB RAM, and running Microsoft Windows 10 Pro. The number of constraints in MILP-Cov and MILP-Mov only differ by one, however MILP-Cov has a consistently lower CPU time than MILP-Mov. This is because, in MILP-Mov, the objective function involves binary variables and the complexity of ILP increases exponentially with integer/binary variables.
%
\begin{table*}
\caption{Comparison of the number of constraints and variables.}
\label{complexity_table}
\centering
\begin{tabular}{cccc}
\hline
Method & No. of Binary Variables & No. of Continuous Variables & No. of Constraints\\
\hline
MILP-Cov &  $LK_{\text{max}}|\mathcal{C}|$ & $(1 + LK_{\text{max}})|\mathcal{C}|$ & $LK_{\text{max}}(3|\mathcal{C}| + 1)$ + $|\mathcal{C}|(2-L)$\\
MILP-Mov & $LK_{\text{max}}|\mathcal{C}|$ & $(1 + LK_{\text{max}})|\mathcal{C}|$ & $LK_{\text{max}}(3|\mathcal{C}| + 1)$ + $|\mathcal{C}|(2-L)+1$\\
MILP-Static & $N_{s}|\mathcal{C}|$ & $N_{s}|\mathcal{C}|$ & $N_{s} + (N_{s} + 1)|\mathcal{C}|$\\
\hline
\end{tabular}
\end{table*}
\begin{table}
\setlength{\tabcolsep}{4pt} 
\renewcommand{\arraystretch}{1.1} 
\caption{Comparison of the number of movements and CPU times required for full coverage by MILP-Cov and MILP-Mov.}
\label{Movements v/s Iterations}
\centering
\begin{tabular}{|c|c|c|c|c|c|c|}
\hline
Network & \multirow{2}{*}{$L$} & \multirow{2}{*}{$N_{s}$} & \multicolumn{2}{c|}{Movements} & \multicolumn{2}{c|}{CPU Time (s)}\\
\cline{4-7}
Size  &       &       &  MILP-Cov & MILP-Mov  &  MILP-Cov & MILP-Mov \\
\hline
\multirow{3}{*}{$10\times 10$} & \multirow{3}{*}{3} & 0  & 17 & 16 &  7.2 & 1573.6\\
        &   & 5  & 12 & 11  & 1.3 & 6.4\\
        &   & 10 & 8 & 6  & 1.5 & 2.3\\
\hline
\multirow{3}{*}{$12\times 12$} & \multirow{3}{*}{3} & 0 & 32 & 32  & 454.2 & 18000\\
        &   & 5  & 18 & 16 & 16.9 & 18000\\
        &   & 10 & 9 & 8  & 3.4 & 117.2\\
\hline
\multirow{3}{*}{$10\times 10$} & \multirow{3}{*}{4} & 0  & 16  & 16  &  3.6 & 1536.8 \\
        &   & 5  & 12 & 11  & 1.2 & 31.9\\
        &   & 10 & 7  & 6  & 1.4 & 2.9\\
\hline
\multirow{3}{*}{$12\times 12$} & \multirow{3}{*}{4} & 0  & 28 & 25  & 137.6 & 18000\\
        &   & 5  & 16 & 16  & 9.0 & 18000\\
        &   & 10 & 10 & 7  &  3.6 & 164.8.1\\
\hline
\multirow{3}{*}{$10\times 10$} & \multirow{3}{*}{5} & 0  & 16  & 16  &  5.6 & 3326.7\\
        &   & 5  & 11 & 11  & 1.3 & 188.4\\
        &   & 10 & 8 & 6   & 1.4 & 3.2\\
\hline
\multirow{3}{*}{$12\times 12$} & \multirow{3}{*}{5} & 0  & 25 & 24  &  118.3 & 18000\\
        &   & 5  & 18 & 16   &  10.4 & 18000\\
        &   & 10 & 10 & 7  & 3.4 & 15.6\\
\hline
\end{tabular}
\end{table}
%
\section{Conclusion}
\label{sec:conclusion}
We have proposed three MILP-based formulations, MILP-Static, MILP-Cov, and MILP-Mov, for efficient placement of static nodes and path planning of mobile nodes in order to maximize the network area coverage and minimize the number of mobile node movements to achieve a desired area coverage. The static node placement strategy addresses the issues that can arise due to random deployment of static nodes and the path planning strategies allow for an explicit limit on overlapping/redundant coverage. The proposed path planning methods achieve improved area coverage and improve the sweep coverage time by minimizing the number of movements, which in turn can improve the network lifetime.
%
\bibliographystyle{IEEEtran}
\bibliography{references}
\end{document}